\documentclass[12pt]{iopart}

\def\sqrtsNN{\mbox{$\sqrt{s_\mathrm{NN}}$}}
\usepackage{graphicx}
\begin{document}

\title[STAR Non-Photonic Electron-Hadron Correlations]{STAR Measurements of Bottom to Charm Ratio and Heavy Quark Interaction with the QCD Medium through Non-Photonic Electron-Hadron Correlations}

\author{Wenqin Xu (for the STAR Collaboration)}

\address{University of California,
Los Angeles, California 90095, USA}
\ead{xwq1985@physics.ucla.edu}

\begin{abstract}
We present STAR measurements of relative charm and bottom contributions to NPE from $p+p$ collisions at $\sqrt{s}= 200$ and 500~GeV energies. We report the total bottom quark production cross section from p+p collisions at $\sqrt{s}= 200$~GeV extracted from NPE spectrum and B to D ratios. We also present the NPE-hadron azimuthal correlations from $Au+Au$ collisions at $\sqrtsNN = 200$~GeV from the 2010 RHIC run where we have collected high statistics data set with low photonic conversion background. 
\end{abstract}


\section{Introduction}
Heavy Quarks are mostly produced through gluon fusions during the initial stage of the heavy ion collisions~\cite{Gyulassy95a}. Experimentally heavy quarks are found to suffer a considerable energy loss in the QCD medium with the nuclear modification factor for non-photonic electrons (NPEs) from heavy quark decays much smaller than unity in central $Au+Au$ collisions at RHIC~\cite{PHENIX10a}. To better understand the heavy flavor production and energy loss mechanism, 
it is crucial to determine experimentally the relative contributions of charm and bottom hadron decays to NPE and to study detailed characteristics of heavy quark interactions with the bulk QCD medium. 

The separation of bottom and charm contributions to NPE has been achieved for $p+p$ collisions at $\sqrt{s}= 200$~GeV by studying the near side of NPE-hadron azimuthal correlation~\cite{STAR10a}. In this work, we extend this analysis to $p+p$ collisions at $\sqrt{s}= 500$~GeV. Since the high $p_T$ ($p_T>3$~GeV/c) NPEs represent well the directions of their heavy flavor parents, and since heavy flavor quarks are produced close to back-to-back, the away side of the NPE-hadron azimuthal correlation in $Au+Au$ collisions bears the information of the processes of heavy flavor quarks traversing bulk QCD medium. As a result, one can use the NPE-hadron azimuthal correlations to study the heavy flavor tagged jet-medium interactions, as we do in this work, which can shed light on the medium responses beyond flow-like correlations.

\section{Analysis}
The high $p_T$ NPE results reported here are based on high-tower triggered events collected by STAR in several RHIC runs. 
The main detectors used are the STAR TPC (Time Projection Chamber)~\cite{STAR03a}, BEMC (Barrel Electromagnetic Calorimeter) and BSMD (Barrel Shower Maximum Detector)~\cite{STAR03b}. 
Electron identification criteria are applied to those tracks that triggered the events. 
The sample of selected electrons, 
referred to as inclusive electrons, includes NPEs,  photonic electrons and the hadron contamination. To estimate the contribution of photonic electrons, the electrons from gamma conversions and $\pi^{0}$, $\eta$ Dalitz decays, we reconstruct the 2D invariant masses of electron pairs, 
and the reconstruction efficiency $\epsilon$ is obtained from embedding. For more details for the analysis methods, see~\cite{STAR10a}. The NPE contribution is statistically obtained by:
\begin{equation}
\Delta \phi_{NPE}=\Delta \phi_{inclusive} - (\Delta \phi_{oppo-sign} - \Delta \phi_{same-sign})/\epsilon - \Delta \phi_{hadron}
\end{equation}
The hadron contribution is neglected, when there are less than 1\% hadrons in the inclusive electron sample; otherwise it is estimated and subtracted.

\section{Results}
Recently, we discovered an error in previous measurements of NPE spectra based on Run03 and Run04~\cite{STAR07a}. An Erratum was published with corrected results~\cite{STAR11b}. Meanwhile, we measured high $p_T$ NPE spectra in $p+p$ collisions at $\sqrt{s}= 200$~GeV based on the data collected in Run05 and Run08. The photonic electron background in Run08 is dramatically lower than that in Run05 due to lower material budget and the independently analyzed NPE spectra in Run05 and Run08 agree with each other. Combining Run05 and Run08 together, we reported a high precision measurement of high $p_T$ NPE spectrum~\cite{STAR11a}, which is consistent with both FONLL prediction~\cite{Matteo05a} and the updated Run03 result in~\cite{STAR11b}.

In the previous analysis of NPE-hadron azimuthal correlation in $p+p$ collisions at $\sqrt{s}= 200$~GeV~\cite{STAR10a}, the relative bottom contribution to NPE was found to be important at high $p_T$. We applied this to the new measurement of NPE spectrum, with quakonium decay contribution subtracted, and obtained the separated spectra for bottom decay electrons and charm decay electrons~\cite{STAR11a}. Based on PYTHIA~\cite{PYTHIA} calculations of bottom decay electrons, which are normalized to the STAR measurements at mid-rapidity at high $p_T$, we extrapolated the total bottom production cross section in $p+p$ collisions at $\sqrt{s}= 200$~GeV to be between $1.34\mu$b and $1.83\mu$b, consistent with FONLL prediction~\cite{Matteo05a}. There are additional experimental uncertainties, i.e. 12.5\%(stat.) and 27.5\%(sys.). 

\begin{figure}[tb]
\centering
\includegraphics[width=0.85 \linewidth]{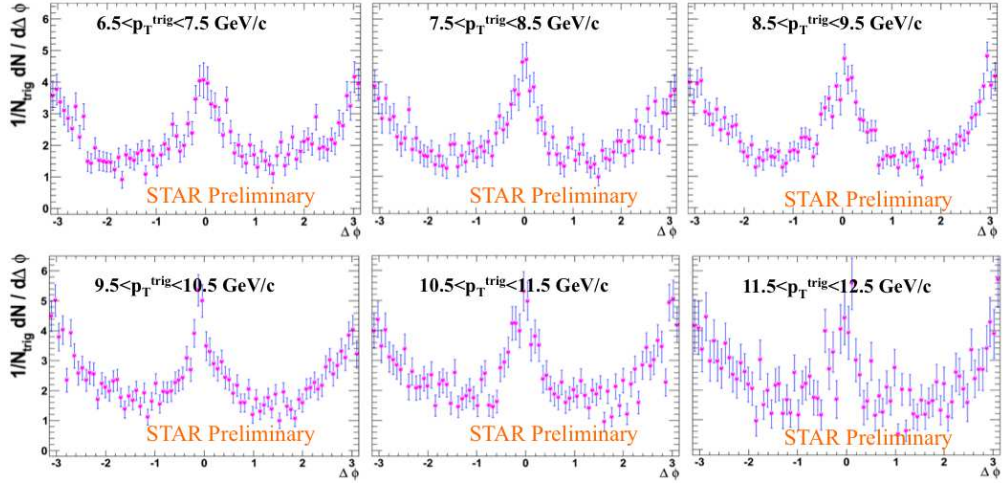}
\caption{NPE-hadron azimuthal correlations in $p+p$ collisions at $\sqrt{s}= 500$~GeV. Different panels are for different NPE $p_T$ ranges.}
\label{fig:NPEh500}
\end{figure}

\begin{figure}[tb]
 \begin{tabular}{cc}
 \includegraphics[width=0.5 \linewidth]{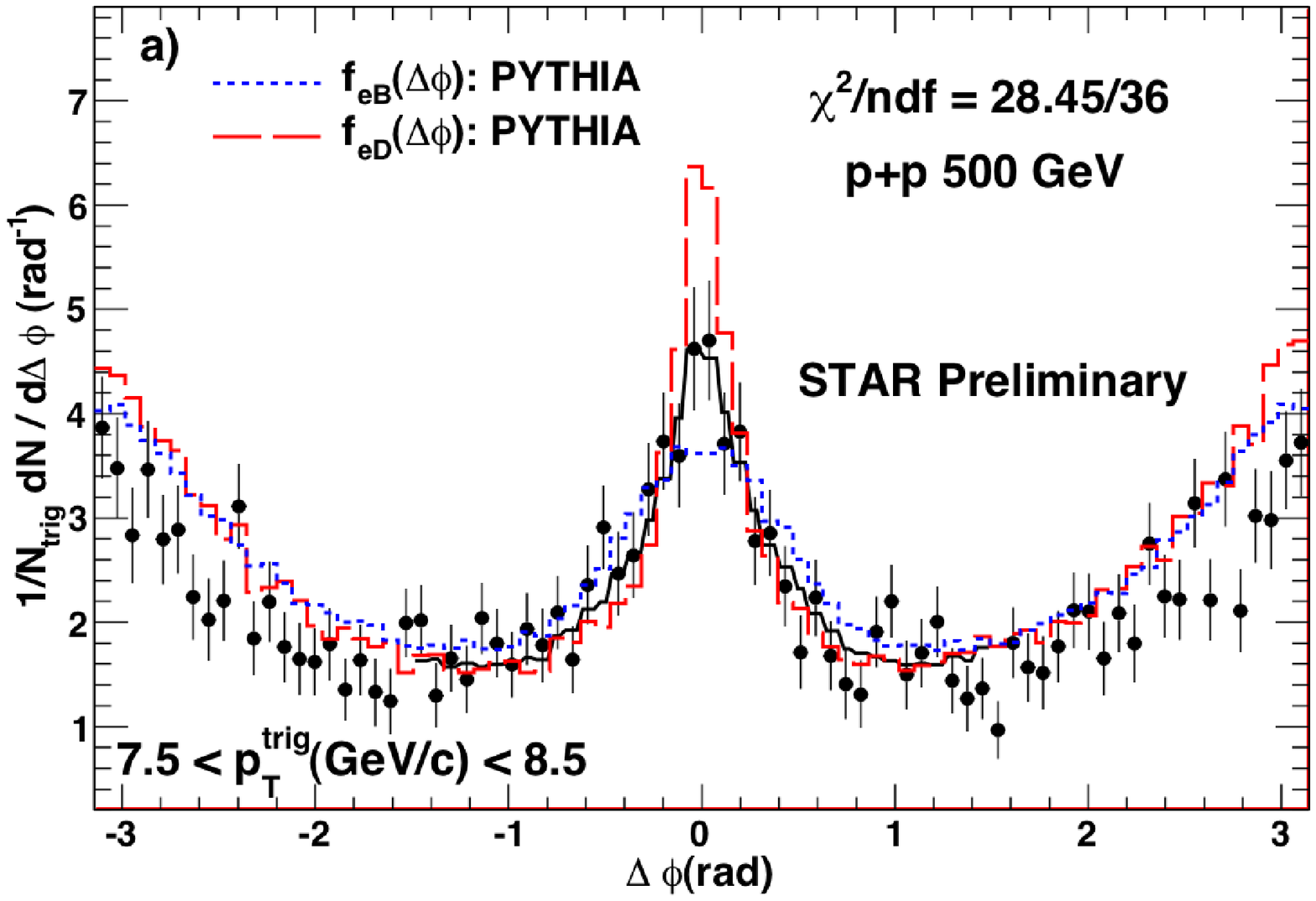}
 \includegraphics[width=0.475 \linewidth]{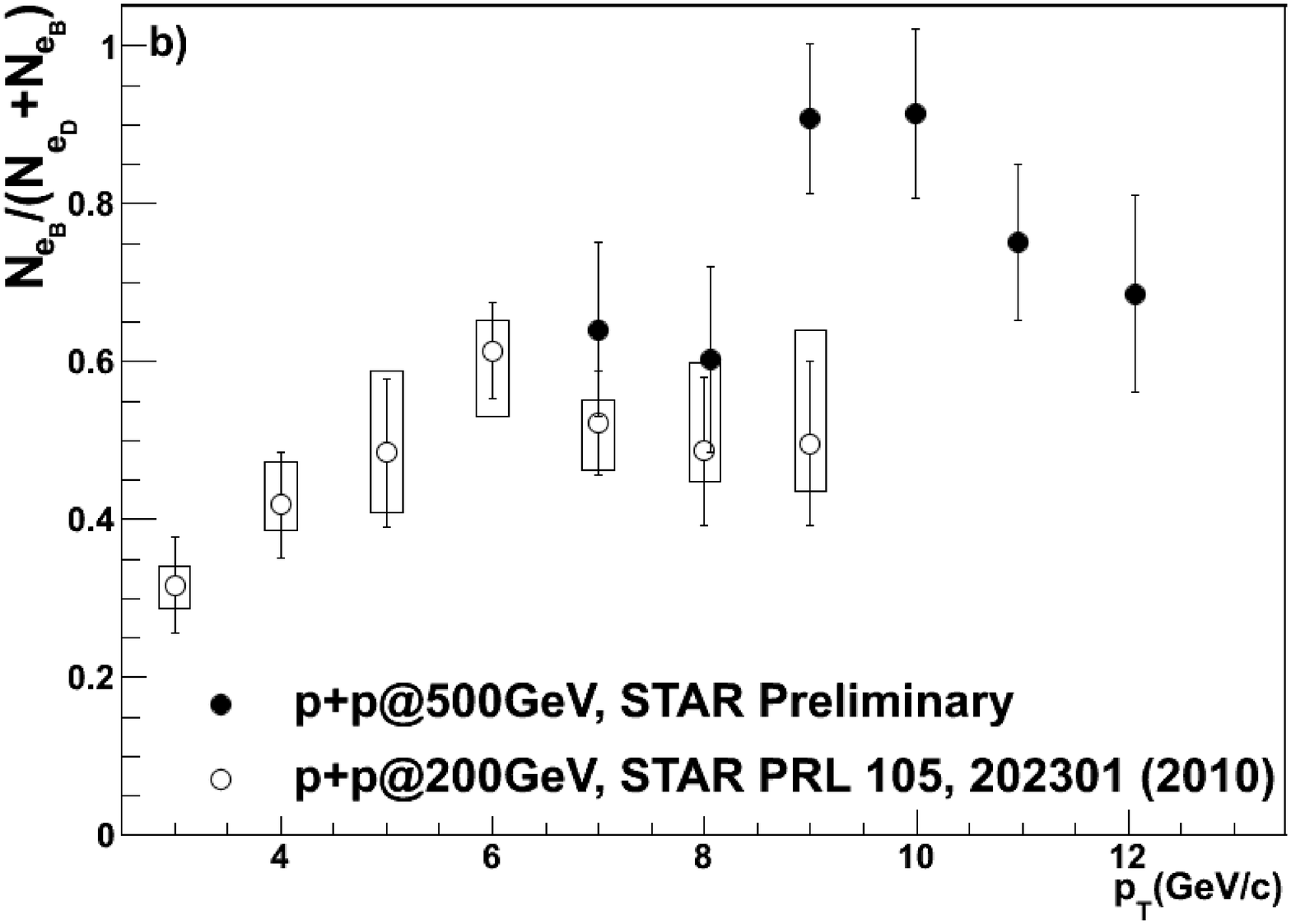}
 \end{tabular}
\caption{Panel (a): NPE-hadron azimuthal correlations from data and PYTHIA. The blue dotted (red dashed) curve is a PYTHIA calculation of B (D) decays and the combination of these two, the black solid curve, is used to fit  the data. Panel (b): The preliminary result of relative bottom contribution to NPE in $p+p$ collisions at $\sqrt{s}= 500$~GeV shown as solid circles with statistical error bars only, and the published result for $200$~GeV shown as open circles with both statistical and systematical errors.}
\label{fig:B500}
\end{figure}

In Figure~\ref{fig:NPEh500}, we show the NPE-hadron azimuthal correlations in $p+p$ collisions at $\sqrt{s}= 500$~GeV, based on high-tower events with an effective threshold of $E_T>7.4$~GeV collected in Run09. The associated tracks have $p_{T}^{asso}>0.3$~GeV/c and $|\eta^{asso}|<1$. In all NPE $p_T$ bins, clear near side correlations are observed. In Figure~\ref{fig:B500}, we compare the near side correlations against corresponding PYTHIA~\cite{PYTHIA8} calculations (Panel a) and extract the relative bottom contribution to NPE (Panel b). The 200~GeV result is also plotted as a comparison, which is  within the range predicted by FONLL calculations~\cite{{Matteo05a},{STAR10a}}. 

\begin{figure}[tb]
\centering
\includegraphics[width=0.65\linewidth]{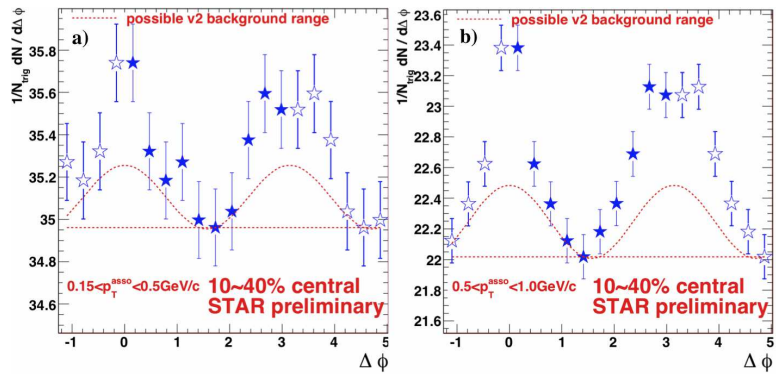}
\caption{Non-photonic electron-hadron azimuthal correlations in $Au+Au$ collisions at $\sqrtsNN = 200$~GeV. The trigger NPE $p_T^{NPE}$ is 3$\sim$6~GeV/c. The $p_{T}^{asso}$ for associated tracks is 0.15$<p_{T}^{asso}<$0.5~GeV/c in Panel (a) and 0.5$<p_{T}^{asso}<$1~GeV/c in Panel (b). $|\eta^{NPE}|<0.7$ and $|\eta^{asso}|<1$ is used. The error bars are statistical only. The correlations are the raw correlations without any background subtraction. The two dotted red curves in each panel represent the maximum and minimum possible $v_2$ background.}
\label{fig:NPEhAuAu}
\end{figure}

Figure~\ref{fig:NPEhAuAu} shows the NPE-hadron azimuthal correlations in 10$\sim$40\% central events of $Au+Au$ collisions at $\sqrtsNN = 200$~GeV with associated tracks at different $p_{T}^{asso}$. The lower $p_{T}^{asso}$ tracks used in panel (a) can better represent the bulk medium, but the statistical error bars are large. With higher $p_{T}^{asso}$, finite correlations beyond the statistical uncertainty are seen, in panel (b), at both near side and away side. However, the elliptic flow ($v_2$) of NPE is still under study, so the $v_2$ background is not subtracted from the raw correlations. To estimate the maximum possible $v_2$ background,  we assume NPE $v_2$  to be the same as hadron $v_2$, with values from Run04~\cite{STAR04a}. The minimum possible $v_2$ background is zero. In each panel, we draw two dotted red curves to show these two limits at positions determined with the Zero Yield at Minimum (ZYAM) approach~\cite{STAR05a}. In additional to the $v_2$ background, the possible flows of other harmonic orders complicate the picture of background composition, requiring further investigations. This study is based on about half of the total high-tower statistics collected in Run10, and the other half is being analyzed. More statistics are also forthcoming in Run11.

In summary, STAR measured high $p_T$ NPE spectra in $p+p$ collisions at $\sqrt{s}= 200$~GeV with higher statistics~\cite{STAR11a}. We presented preliminary results of relative bottom contribution to NPE at high $p_T$ in $p+p$ collisions at $\sqrt{s}= 500$~GeV and NPE-hadron azimuthal correlations in $Au+Au$ collisions at $\sqrtsNN = 200$~GeV.

\end{document}